\documentclass[conference]{IEEEtran}
\IEEEoverridecommandlockouts
\pdfoutput=1
\usepackage{cite}
\usepackage{amsmath,amssymb,amsfonts}
\usepackage{booktabs}
\usepackage{pifont}
\usepackage[usenames, dvipsnames]{color}

\usepackage[table,xcdraw]{xcolor}
\usepackage{dirtytalk}
\usepackage{multirow}
\usepackage[table,xcdraw]{xcolor}
\usepackage{lipsum}
\usepackage{subfig}
\usepackage{blindtext, graphicx}
\usepackage{amsmath}
\usepackage{mathtools}
\usepackage{xspace}
\newcommand{\proposedapp}{MAPS\xspace}

\usepackage{multicol}
\usepackage{graphicx}
\usepackage{fullpage}
\usepackage{times}
\usepackage{fancyhdr,graphicx,amsmath,amssymb}

\usepackage[ruled,vlined,linesnumbered]{algorithm2e}

\usepackage{balance}

\usepackage{inputenc}
\graphicspath{{Pictures/}}

\usepackage{xcolor}

\usepackage{setspace}
\usepackage[top=1in,bottom=1in,right=0.75in,left=0.75in,headheight=65pt]{geometry}
\usepackage{fancyhdr}
\def\BibTeX{{\rm B\kern-.05em{\sc i\kern-.025em b}\kern-.08em
    T\kern-.1667em\lower.7ex\hbox{E}\kern-.125emX}}
    
\fancypagestyle{firstpage}{%
 \fancyhf{}%
 \fancyhead[C]{2024 5th CPSSI International Symposium on Cyber-Physical Systems (Applications and Theory) (CPSAT)}
  
 \fancyfoot[L]{%
  \normalsize{979-8-3315-2928-4/24/\$31.00~\copyright2024 IEEE}
 }%
}

\ifCLASSINFOpdf
\else
\fi
\makeatletter

\def\ps@IEEEtitlepagestyle{
  \def\@oddfoot{\mycopyrightnotice}
  \def\@evenfoot{}
}

\def\mycopyrightnotice{
  {\footnotesize
  979-8-3315-2928-4/24/\$31.00~\copyright2024 IEEE\hfill} % <--- Change here
  \gdef\mycopyrightnotice{}
}

\makeatother
\usepackage[]{siunitx}
\usepackage{textcomp}
\begin{document}
\include{pythonlisting}
\title{\proposedapp: Energy-Reliability Tradeoff \underline{M}anagement in \underline{A}utonomous Vehicles Through LLMs \underline{P}enetrated \underline{S}cience}
%\conf{2020 CSI/CPSSI International Symposium on Real-Time and Embedded Systems and Technologies (RTEST), June 10-11, 2020, Institute for Research in Fundamental Sciences (IPM), Tehran, Iran.}

% \author{\IEEEauthorblockN{Mahdieh Aliazam\IEEEauthorrefmark{1}, Ali Javadi\IEEEauthorrefmark{2},
% Amir Mahdi Hosseini Monazzah\IEEEauthorrefmark{3} and Ahmad Akbari Azirani\IEEEauthorrefmark{4}}

% \IEEEauthorblockA{\IEEEauthorrefmark{1}\IEEEauthorrefmark{2}\IEEEauthorrefmark{3},\IEEEauthorrefmark{4}School of Computer Engineering, Iran University of Science and Technology, Tehran, Iran\\
% Email: {\IEEEauthorrefmark{1}mahdieh\_aliazam77@comp.iust.ac.ir,
% \IEEEauthorrefmark{2}javadi\_ali@comp.iust.ac.ir, \IEEEauthorrefmark{3}monazzah@iust.ac.ir
% \IEEEauthorrefmark{4}akbari@iust.ac.ir}
% }
% }

\author{\IEEEauthorblockN{Mahdieh Aliazam, Ali Javadi, Amir Mahdi Hosseini Monazzah, and Ahmad Akbari Azirani}
\IEEEauthorblockA{\textit{School of Computer Engineering},\\
\textit{Iran University of Science and Technology},\\
Tehran, Iran \\
\{mahdieh\_aliazam77, javadi\_ali\}@comp.iust.ac.ir, \{monazzah, akbari\}@iust.ac.ir}
}

\maketitle
\thispagestyle{firstpage}  % Apply fancy header on the first page
%\IEEEpubidadjcol

\begin{abstract}
As autonomous vehicles become more prevalent, highly accurate and efficient systems are increasingly critical to improve safety, performance, and energy consumption. Efficient management of energy-reliability tradeoffs in these systems demands the ability to predict various conditions during vehicle operations. With the promising improvement of Large Language Models (LLMs) and the emergence of well-known models like ChatGPT, unique opportunities for autonomous vehicle-related predictions have been provided in recent years. This paper proposed \proposedapp using LLMs as map reader co-drivers to predict the vital parameters to set during the autonomous vehicle operation to balance the energy-reliability tradeoff. The \proposedapp method demonstrates a 20\% improvement in navigation accuracy compared to the best baseline method. \proposedapp also shows 11\% energy savings in computational units and up to 54\% in both mechanical and computational units.

\end{abstract}

\IEEEpeerreviewmaketitle

\section{Introduction}
\label{intro}
\emph{Autonomous vehicles} (AVs), with their advanced technology, will play a significant role in the future of transportation and our daily lives. Modern automobiles increasingly utilize sensors, Advanced Driver Assistance Systems (ADAS), and safety features, moving towards full autonomy \cite{SurveyOnAV}. These technologies include integrating sensors with advanced deep-learning models that can assist or replace the driver. Full autonomy means performing driving processes from start to finish without human intervention. Integrating cameras, LiDAR, global navigation systems, radar, and communication modules with advanced software enables automated driving. Features such as brake assist, lane departure warning, and adaptive cruise control, introduced in the 1990s, have somewhat improved the safety of vehicles on the roads \cite{yeong2021sensor}.

The current trend involves integrating deep learning and machine learning methods into AVs to achieve maximum reliability (accuracy). These learning algorithms aim to interpret the driving environment, receive various environmental data as input, and provide outputs based on input features. Machine learning algorithms have created a significant transformation in the development of AVs. By processing vast amounts of data in real time, these algorithms allow AVs to make complex decisions quickly and accurately. For instance, neural networks can analyze car camera images to identify objects, detect road lines, and determine the driving path \cite{bathla2022autonomous}.

While the technological advancements of AVs bring numerous benefits, it is essential to consider the energy consumption associated with these systems \cite{peterautonomous}. AVs rely on a suite of sensors, computational resources, and mechanical components to function effectively. These components, including high-resolution cameras, LiDAR, powerful onboard processors for complex algorithm processing, and motors, consume substantial energy. With the increasing demand for AVs, energy consumption could become a serious concern, affecting operational costs and reliability \cite{SurveyOnAV}.

Reducing energy consumption in AVs is crucial for several reasons. Lower energy consumption can increase the traveled distance of electric AVs, making them more practical for long-distance travel and reducing the need for recharging \cite{EnergyEfficientMobile}. On the other hand, the accuracy and reliability of these vehicles are also of high importance, as high accuracy can prevent accidents and ensure passenger safety. Furthermore, greater reliability and compatibility of the vehicle in various road conditions can reduce energy consumption and operational costs \cite{lu2021real}.

One way to reduce energy consumption while considering accuracy and reliability is to use Large Language Models (LLM) as map reader co-drivers to predict critical parameters for adjustment during autonomous vehicle operations, thereby balancing energy and reliability consumption. LLM are advanced AI systems designed to understand and generate human-like text based on vast amounts of data. These models, such as OpenAI's GPT \textsuperscript{\textregistered}, use deep learning techniques, particularly transformer architectures, to process and analyze text. LLM can perform a wide range of natural language processing tasks by being trained on extensive and diverse datasets \cite{chang2024survey}.

In the context of AVs, LLM can play a significant role in enhancing performance and optimizing accuracy and energy consumption. By leveraging LLM with vehicle systems as map reader co-drivers and predicting critical environmental parameters, more precise decisions can be made for proper navigation with appropriate speed and accuracy. This can lead to smoother driving patterns, reducing unnecessary acceleration and braking and saving energy \cite{kong2024superalignment}.

Given the above motivation, in this paper, we present \proposedapp. \proposedapp is an efficient LLM strategy to create a tradeoff between energy consumption and reliability. \proposedapp aims to increase routing accuracy as a reliability metric and manage energy consumption by predicting route challenges using LLM and utilizing LLM to manage mechanical and computational actuators. In \proposedapp, we used ChatGPT \textsuperscript{\textregistered} as a map reader co-driver to test the proposed method. We implemented our experiments in a real environment using a Raspberry Pi board and DC motors on a \proposedapp-equipped autonomous robot car. Overall, the contributions of \proposedapp can be summarized as follows:

\begin{itemize}
    \item Using LLM as a map reader co-drivers 
    
    \item Predicting critical environmental parameters and making decisions based on them.
    
    \item Control policy for selecting the best motor speed and processing accuracy.

    \item Practical testing demonstrating \proposedapp increased routing accuracy and energy savings compared to the baseline method
\end{itemize}

We evaluate the efficiency of \proposedapp and compare the evaluation results with baseline scenarios. Our experiments show that the \proposedapp method has achieved a 20\% improvement in navigation accuracy compared to the best baseline method with high speed and Frames Per Second (FPS). It also shows 11\% more energy efficiency per computational unit compared to baseline methods with the highest energy consumption and up to 54\% total energy savings compared to other baseline methods with high speed and FPS.

The rest of the paper is organized as follows. Section II reviews the background of AVs and LLM. Then, we examine related work in Section III. In Section IV, the \proposedapp approach is explained. Simulation results are discussed in Section V. Finally, we conclude the paper in Section VI. Selected prompts and responses of the LLM can be found in the Appendix.

\section{Background}
\label{preliminaries}
This chapter addresses the technology of AVs and analyzes the significant role of energy consumption in these vehicles. An introduction to LLM and their applications in AVs will be discussed. These topics aim to provide a comprehensive overview of the technological advancements and challenges associated with AVs and the role of LLM in enhancing their performance and accuracy.

\subsection{Autonomous Vehicle Technology}

The technology of AVs comprises several key components, including sensors, algorithms, and control systems. Sensors are responsible for monitoring, obstacle detection, and navigation. These include cameras, LiDAR, radar, and ultrasonic sensors. They enable the vehicle to accurately identify its surroundings and provide the necessary data for real-time and precise decision-making \cite{peterautonomous}.

Algorithms are another crucial part of autonomous vehicle technology. They process the data received from sensors and make the necessary decisions to guide the vehicle. These algorithms include artificial neural networks, machine learning, and decision-making algorithms that analyze environmental data select optimal paths, and respond appropriately to various situations. Advanced algorithms can accurately manage complex behaviors such as lane changes, stopping at red lights, and collision avoidance \cite{bachute2021autonomous}.

Control systems, another vital component of this technology, execute the commands issued by the algorithms. These systems include electronic and mechanical controllers that directly interact with the physical components of the vehicle, such as the steering, brakes, and throttle. Control systems must execute high-precision and speed commands to ensure the vehicle moves safely and efficiently. Their role in maintaining the stability and accuracy of autonomous vehicle operations is critical \cite{yao2020control}.

Integrating software and hardware in AVs is also of particular importance. This integration involves the complete coordination between sensors, algorithms, and control systems to enable the vehicle to operate seamlessly and cohesively. Advanced software must rapidly process various data and relay critical information to control systems in real time. Additionally, the hardware must exhibit high precision and stability to execute software commands accurately \cite{collin2020autonomous}.

Ultimately, the success of autonomous vehicle technology depends on the coordinated and effective interaction between all these components. Each part must function flawlessly to ensure the vehicle moves safely and efficiently. Therefore, continuous development and improvement in sensors, algorithms, and control systems, along with integrating software and hardware, are essential for advancing and broader adoption of AVs \cite{peterautonomous}.

\subsection{Energy Consumption in AVs}

\begin{figure}[t]
	\centering
 \includegraphics[ width=0.45\textwidth]{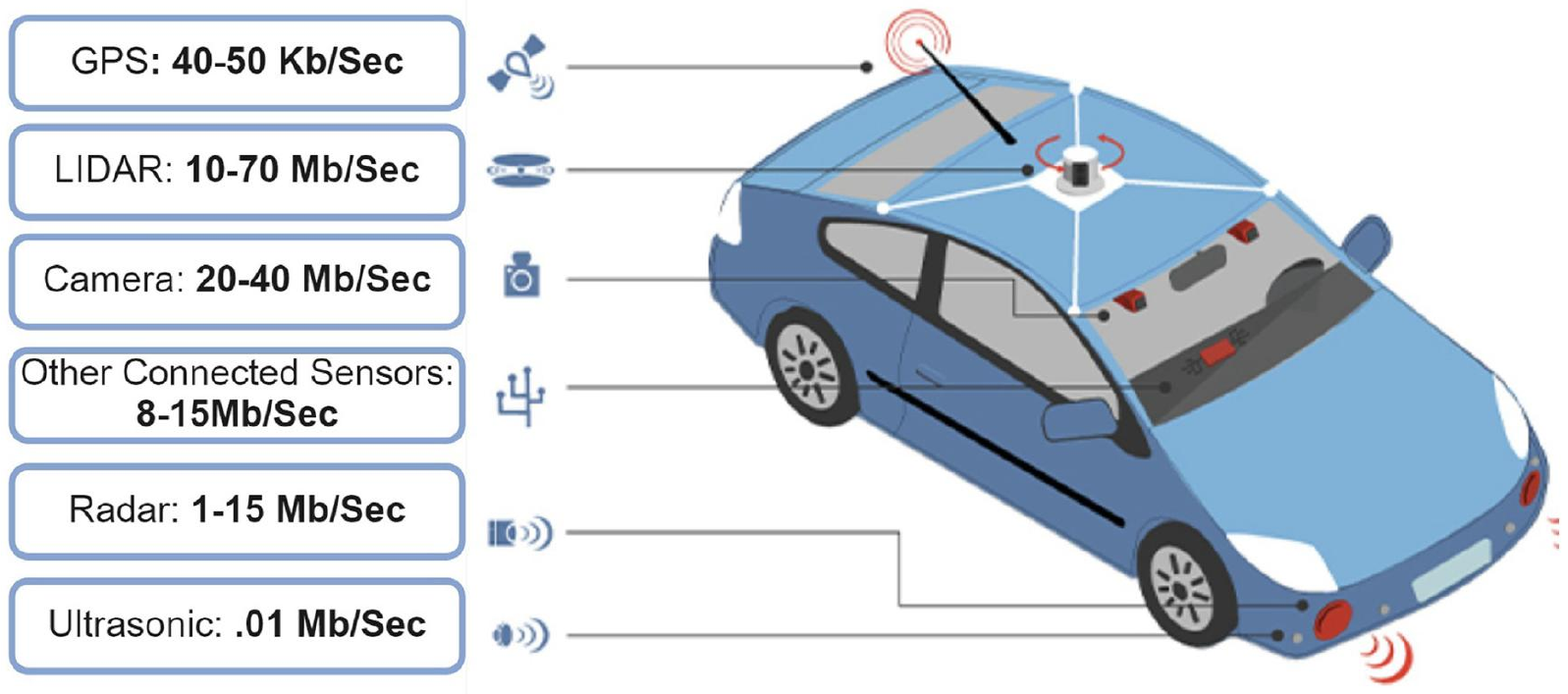}
	\caption{Data generated by the automotive sensors \cite{SurveyOnAV}}
	\label{fig:AVsData}
\end{figure}

In AVs, sensors such as cameras, LiDAR, and radar generate large amounts of raw data that the vehicle's computing unit must process. As shown in Fig \ref{fig:AVsData}, the data rate from these sensors varies depending on their technical specifications, such as generation, bit rate, and recording features \cite{SurveyOnAV}. For example, the data rate produced by a LiDAR sensor may differ from that of a camera, as each sensor collects different types of data from the vehicle's surroundings that need processing.

These factors directly impact the energy consumption of AVs. Generally, energy consumption in AVs can be divided into three main categories. The first category is energy consumption by the vehicle's sensors, computing devices, and mechanical components, which accounts for most energy use. The second category is energy consumption due to infrastructure sensors and vehicular network communications, essential for coordination and information exchange between vehicles and infrastructure \cite{malawade2022ecofusion}. The third category pertains to energy consumption in the backend, such as Edge servers and local and central servers, which store and process historical data. The level of autonomy in AVs significantly affects energy consumption, as higher levels of autonomy require more sensors, computing units, and controllers \cite{lu2021real}.

Vehicle autonomy is divided into six levels, each requiring a specific set of sensors and requirements. At Level 0, there is no automation, and driving is entirely dependent on the driver. Level 1 includes driver assistance, where driving tasks are carried out with the help of inputs from vehicle sensors. At Level 2, partial automation exists, and some driving tasks, such as adaptive cruise control and emergency braking, are performed by the vehicle's computing unit. However, the driver must still maintain control.

At Level 3, conditional automation allows the vehicle to perform some tasks autonomously, but the driver must take control when necessary. Level 4 offers high automation, where the vehicle can perform all driving tasks under certain conditions, but the driver can still take control if needed. Finally, Level 5 includes full automation, where the vehicle can perform all driving tasks in all conditions, although the driver may still have the option to control the vehicle \cite{SurveyOnAV}.

High energy consumption in AVs is due to the use of compute-intensive algorithms and processing devices, such as graphics processors, which are essential for perception and visual applications. One effective approach to reducing energy consumption in AVs is route planning and optimization. This method uses advanced algorithms to determine the best route with optimal speed. By reducing travel time and distance, energy consumption can be significantly decreased. Additionally, using adaptive and predictive models to optimize energy consumption is crucial. These models can analyze past data and predict future needs to optimize vehicle energy consumption \cite{perger2020energy}. For instance, using LLM to predict road conditions and determine optimal speed and accuracy for vehicle movement can be one of these methods \cite{chang2024survey}.

\subsection{Large Language Models (LLM) }

\begin{figure}[t]
	\centering
	\includegraphics[width=\linewidth]{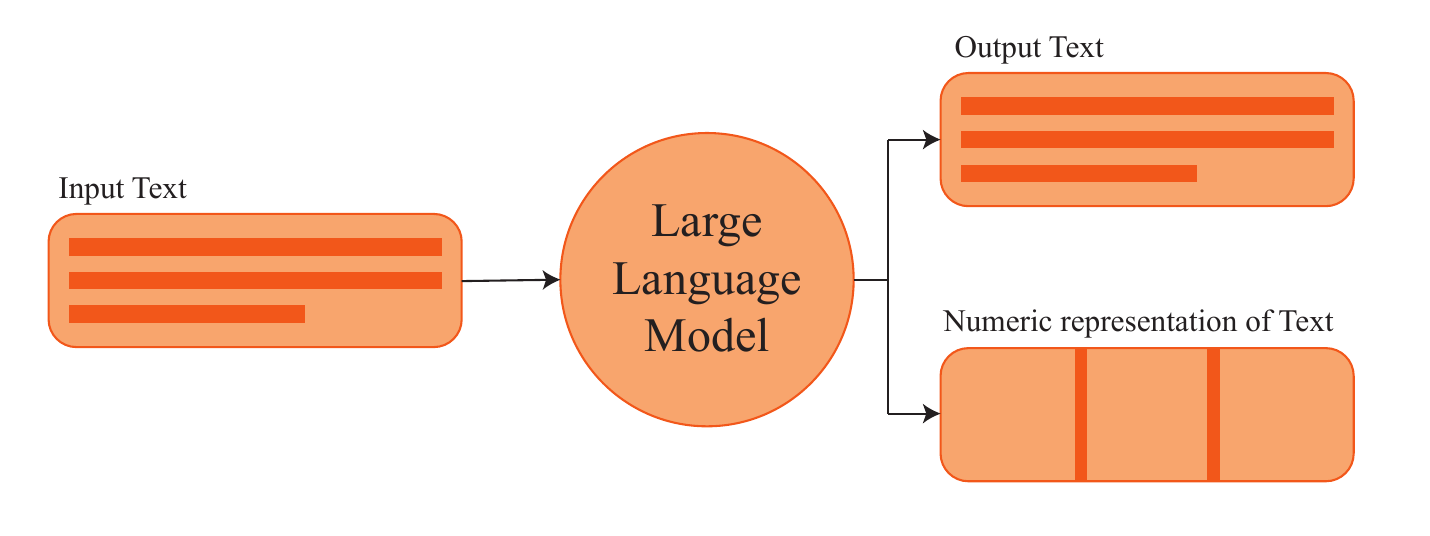}
	\caption{Large Language Models Architecture}
	\label{fig:LLM}
\end{figure}

LLMs are a deep learning model used for processing and generating natural language. These models are trained using large neural networks and vast amounts of text data from various sources to understand and produce human language \cite{ChatGPT}. Well-known models such as GPT-3 and GPT-4 from OpenAI are examples of LLMs, which have been trained with billions of parameters and can perform diverse tasks such as language translation, text generation, and answering questions. These models typically employ advanced machine learning methods like deep learning and Transformers to identify complex patterns and relationships between words and sentences \cite{chang2024survey}.

Fig \ref{fig:LLM} illustrates an LLM Architecture and explains how it processes text inputs and generates text outputs and numerical representations of text. Text input is fed into the LLM on the left side of the image. This model analyzes and processes the input text using deep neural networks and machine learning techniques. This analysis includes understanding the meaning of words and sentences and identifying linguistic patterns, which allows the model to answer questions, generate new texts, or translate content.

The outputs of an LLM are shown in two different forms: text output and numeric representation of text. The text output is the text generated by the model based on the input analysis. This text can be an answer to a question, a new article, or a translation. On the other hand, the numeric representation of text indicates the conversion of text into a series of numbers, which is helpful for machine learning models. These representations help the model better understand complex patterns and relationships between words and sentences, thus increasing the accuracy and efficiency of subsequent processes \cite{huang2023advancing}.

LLMs play a significant role in AVs' map reading and navigation. By analyzing and understanding textual and visual data, these models can accurately and efficiently process complex geographical information. For example, LLMs can analyze digital map data, traffic reports, and weather conditions to suggest the best route for AVs, reducing travel time and increasing efficiency \cite{chang2024survey}.

Moreover, using large language models can significantly enhance the accuracy of autonomous systems. LLMs can analyze various data with high precision and make more accurate decisions by learning from past data. These models can intelligently identify information such as road obstacles, speed limits, and sudden changes in the route, planning appropriate responses. These capabilities ensure that AVs can operate more safely and efficiently, reduce the risk of accidents, and provide a better experience for passengers. Therefore, LLMs play a crucial role in improving the accuracy and efficiency of navigation and autonomous driving systems \cite{kong2024superalignment}.

\section{Related Work}
\label{related}

In this section, we present some studies related to our work that focus on improving routing accuracy in AVs or saving energy in AVs.

In \cite{EnergyEfficientMobile}, the authors investigate and propose an energy optimization controller for mobile robots that uses event-based cameras to perform vision-based operations in real time. This controller simultaneously manages the CPU's voltage/frequency and the mechanical motor voltage to minimize energy consumption. The main idea of this paper is that independently controlling the robot's speed and CPU voltage/frequency does not necessarily lead to an optimal energy solution. To achieve the highest efficiency, computational and mechanical controls must be coordinated. To this end, the paper proposes a fast hill-climbing optimization algorithm that finds the best CPU/motor configuration during runtime when encountering new environments. Experimental results show that the proposed controller can save, on average, 50.5\%, 41\%, and 30\% energy in low, medium, and high complexity environments, respectively, compared to baseline methods. These results were obtained on a robot equipped with brushless DC motors, a Jetson TX2 board as the computational unit, and a DAVIS-346 event-based camera.

In \cite{malawade2022ecofusion} introduces an EcoFusion method for sensor fusion in AVs, aiming to reduce energy consumption without compromising object detection performance. This method dynamically changes the sensor fusion and fusion location based on environmental conditions to optimize energy consumption and detection accuracy. The main idea is that different driving conditions (such as city driving or rain) require different resources and sensors, and identifying these conditions allows for optimal sensor fusion. The researchers aim to optimize energy consumption in object detection systems of AVs by using various sensors (such as cameras, lidar, and radar) and considering environmental conditions. Results show that EcoFusion provides, on average, 9.5\% better object detection performance than existing sensor fusion methods. Additionally, this method reduces energy consumption by approximately 60\% and decreases latency by 58\% compared to the Nvidia Drive PX2 hardware platform.

Wan et al. in \cite{wan2023berry} discusses a framework called BERRY, designed to improve energy efficiency and bit error robustness in reinforcement learning models used in autonomous systems. Reducing the operating voltage can save energy but leads to bit errors that compromise mission safety and performance. BERRY combines robust learning offline and onboard, allowing systems to operate reliably at reduced voltage while achieving significant energy savings. Experimental results show that this framework can reduce energy consumption by up to 15.62\% and increase the number of successful missions by up to 18.51\%.

In \cite{ChatGPT}, researchers examine using LLMs such as ChatGPT as vehicle driving assistants. This research aims to bridge the gap between human intentions, machine understanding, and execution. The study designs a global framework that employs LLMs as a "Co-Pilot" in vehicles to perform specific driving tasks based on human intentions. The performance of this Co-Pilot is improved using a method called black-box tuning. In the experiments, the Co-Pilot was used for two tasks, including path control and route planning. Results showed that the Co-Pilot could perform most tasks using natural language processing, although it is not flawless. Overall, this framework has high potential for broader applications in AVs and can help improve human-machine co-driving.

\section{\proposedapp in Details}
\label{proposed}

\begin{figure}[t]
	\centering
	\includegraphics[width=\linewidth]{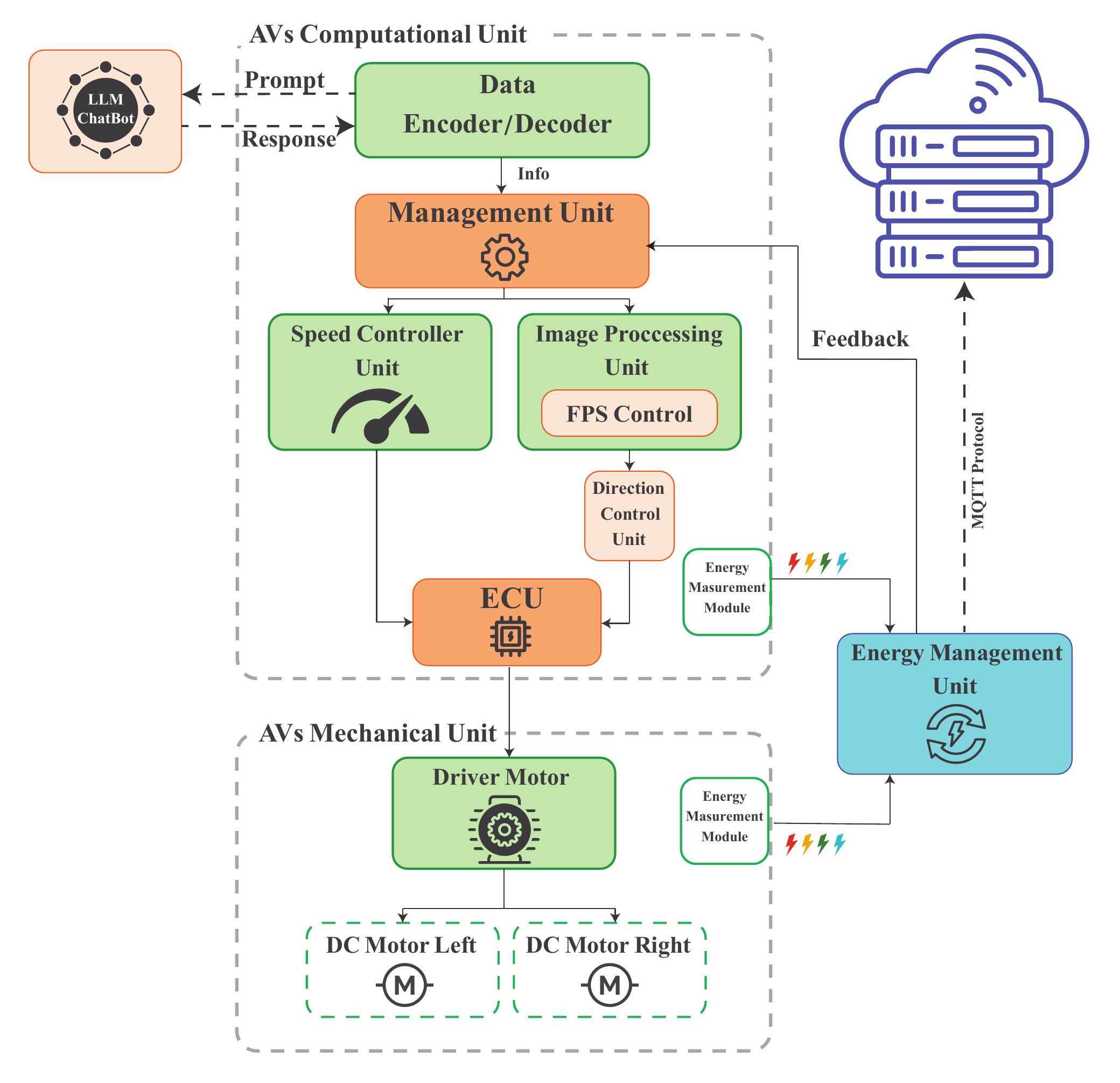}
	\caption{\proposedapp Architecture}
	\label{fig:MAPS}
\end{figure}

As discussed in previous sections, energy consumption and reliability are two critical parameters in AVs. Each of the sensors, computational units, and mechanical components independently significantly impacts the energy consumption of AVs. Independent control of speed (Mechanical Unit) and CPU (Computational Unit) processing in AVs does not necessarily lead to an optimal solution. In other words, merely reducing speed or decreasing computational accuracy does not lead to reduced energy consumption. Rather, to achieve the highest efficiency, computational and mechanical controls must be performed together and in coordination \cite{EnergyEfficientMobile}. Therefore, considering this fact, a solution is needed that can dynamically adjust speed and processing accuracy based on environmental conditions and road changes. This would not only maintain accuracy in path-finding but also significantly save energy, achieving a balance between energy consumption and reliability, which is considered navigation accuracy in this research.

To this end, we present the \proposedapp. \proposedapp uses LLMs chatbot as map reader co-drivers to predict the operational route and adjust critical vehicle parameters such as speed and image processing accuracy to balance accuracy and energy consumption. Fig. \ref{fig:MAPS} shows the \proposedapp architecture. The \proposedapp architecture consists of two parts: the Computational unit and the Mechanical unit. In the Computational unit, relevant parameters, along with an image of the road the vehicle is to travel on, are first converted into a prompt by the Data Encoder/Decoder unit and sent to the chatbot (A detailed example of a prompt can be found in the Appendix). The chatbot, considering the travel environment, the degree of curvature or straightness of the lines, and environmental conditions, sends the relevant response back to the Data Encoder/Decoder unit. Then, the response is sent to the Computational unit of the autonomous vehicle. The data received from the chatbot is decoded, and the necessary information is extracted. 

After extracting the information received from the chatbot, the obtained information, which includes the vehicle's speed and the desired frame rate per second (FPS) for image processing, is provided to the Management unit. Based on the straightness of the path, the chatbot considers a specific speed and FPS, and in the case of a curved path, it considers another value for speed and FPS. Then, in the Management unit, decisions are made, and relevant information is sent to each section. Information regarding FPS is sent to the Image Processing unit, and information regarding speed control is sent to the Speed Control unit. The processed information in the Image Processing unit is applied to the Direction Control unit to change the steering and vehicle path-finding direction.
The Speed Control unit calculates the specified speed and sends it to the Engine Control Unit (ECU). Based on the specified direction and speed, the ECU sends the necessary commands for motor movement to the Driver Motor in the vehicle's Mechanical unit. The Driver Motor decides and activates one of the right or left DC motors or both based on the direction and degree of curvature.

Another unit in the \proposedapp architecture is the Energy Measurement Module, where the energy consumption of the Computational unit and the Mechanical unit (motors) is measured. The measured data of each unit's power consumption is sent to the energy measurement unit, and the power consumption is calculated. The results are also sent and stored on a cloud server using the MQTT protocol. This stored information can be used to improve algorithms and further optimize energy consumption, as well as enable further analysis of the collected data to understand energy consumption patterns better and improve system performance. In \proposedapp, to measure the energy consumption of the Mechanical and Computational units, we calculate their power consumption. First, we calculate the input voltage. The formula \ref{eq:inputV} shows the calculation of the input voltage value.

    \begin{equation}
    \label{eq:inputV}
        \text{InputVoltage}=\frac{\text{Dout} \times \text{Vmax}}
        {\text{Dmax}}
    \end{equation}

In (\ref{eq:inputV}), Dout is the digital output result from the ADC, Vmax is the maximum measurable input analog voltage, and Dmax is the maximum raw digital reading result from the ADC. Then, the current should be calculated. The (\ref{eq:current}) shows the calculation of the current value. The InputVoltage is ADC voltage that calculated in (\ref{eq:inputV}) and the resistance value is the value of shunt resistor.

    \begin{equation}
    \label{eq:current}
       \text{Current} = \frac{\text{InputVoltage}}{\text{Shunt Resistance value}}
    \end{equation}

Finally, the Power consumed by the device is measured with (\ref{eq:power}). In  (\ref{eq:power}), the obtained current is multiplied by the device voltage. This voltage is the voltage of the device we want to measure that power. 

    \begin{equation}
    \label{eq:power}
\text{Power} = \text{Current} \times \text{Device Voltage}
    \end{equation}

Finally, the energy management unit gives the measured energy to the feedback management unit so that new decisions can be made for the next decisions and to control the speed and accuracy of the vehicle.
\section{System Setup and Results}
\label{result}

\begin{figure}[t]
	\centering
	\includegraphics[width=\linewidth]{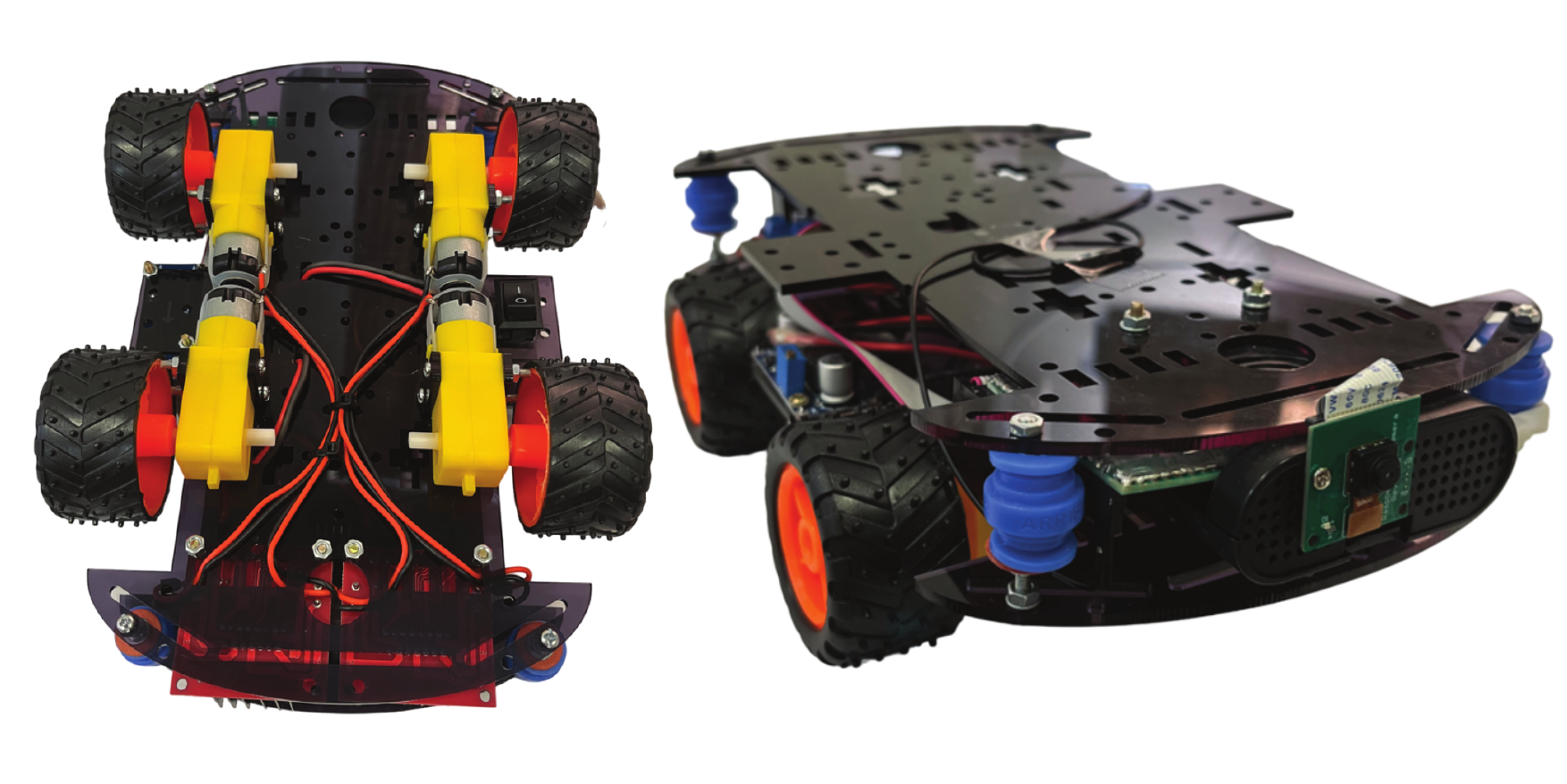}
	\caption{\proposedapp AVs Implemention}
	\label{fig:AVIMP}
\end{figure}

To evaluate the \proposedapp method, we implemented our autonomous vehicle robot in a real-world environment. Our implementation for building and deploying the autonomous vehicle is divided into two sections. The first section is the computational unit, which includes a Raspberry Pi 4B board for processing tasks, code development, and server communication. A Raspberry Pi Camera is also used for video capture and image processing. The second section is the mechanical unit responsible for the robot's movement and speed. The motors are controlled using an L298N motor driver, and DC motors drive the wheels. The implemented robot is shown in Fig. \ref{fig:AVIMP}.

\begin{figure}[t]
	\centering
	\includegraphics[width=\linewidth]{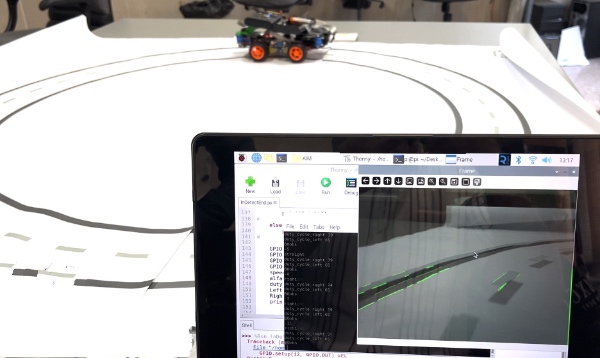}
	\caption{Experiments and Test of AVs}
	\label{fig:Test}
\end{figure}

Python was used to develop the computational unit's code. Lane detection algorithms and the OpenCV library were employed to detect road lines and change directions for navigation. Control commands for the motors are sent via the RPi.GPIO library, which is used to interface with the Raspberry Pi's output pins. Additionally, the wheel rotation speed is controlled using PWM (Pulse Width Modulation), an effective method for controlling the speed of DC motors by varying the pulse width. The information related to road line detection and direction change is received and processed by image processing algorithms, and necessary control actions are sent to the robot's motors. In Fig. \ref{fig:Test}, the conducted experiments and the testing of the self-driving car can be seen. The path for conducting the experiments is a circular route (as seen in the Appendix) where the autonomous vehicle is placed for testing and evaluation. Additionally, the autonomous vehicle has been tested by driving five laps on the test route for each scenario, and the results have been measured based on these tests.

\begin{figure}[t]
	\centering
	\includegraphics[scale=0.15]{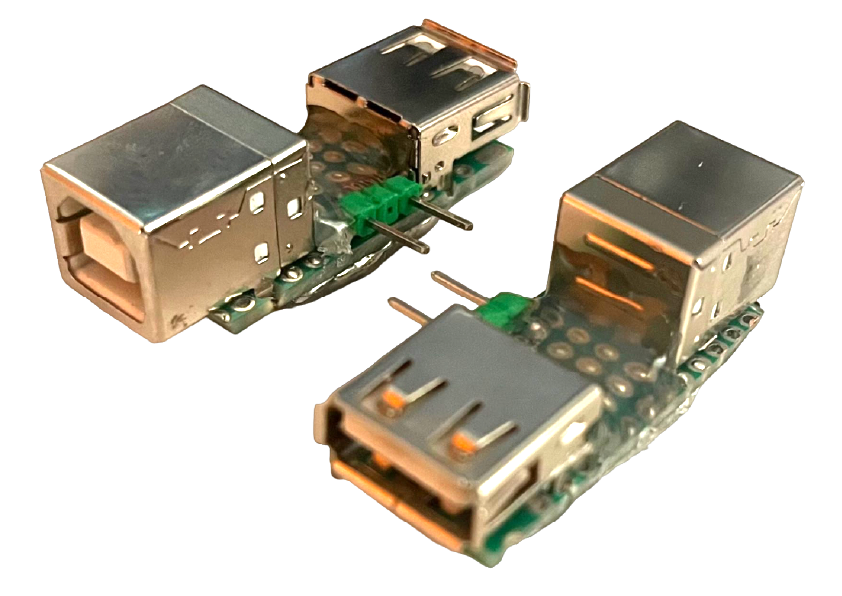}
	\caption{Energy Measurement Module}
	\label{fig:Shant}
\end{figure}

A module using a shunt resistor was designed to measure the energy consumption of each of the computational and mechanical units in the energy management unit, as shown in Fig. \ref{fig:Shant}. This module is placed between the power source and the computational and mechanical units to measure the voltage consumed by each section. The measured values are calculated using formulas \ref{eq:inputV}-\ref{eq:power} from the previous section, and the power consumption of each unit is determined. The measured power has been recorded for the duration of the five test laps. Finally, to determine the final power measurement, we calculate the overall average of the obtained power values. The power consumption data is also sent to the server for storage using the MQTT protocol.

\begin{figure}[t]
	\centering
	\includegraphics[width=\linewidth]{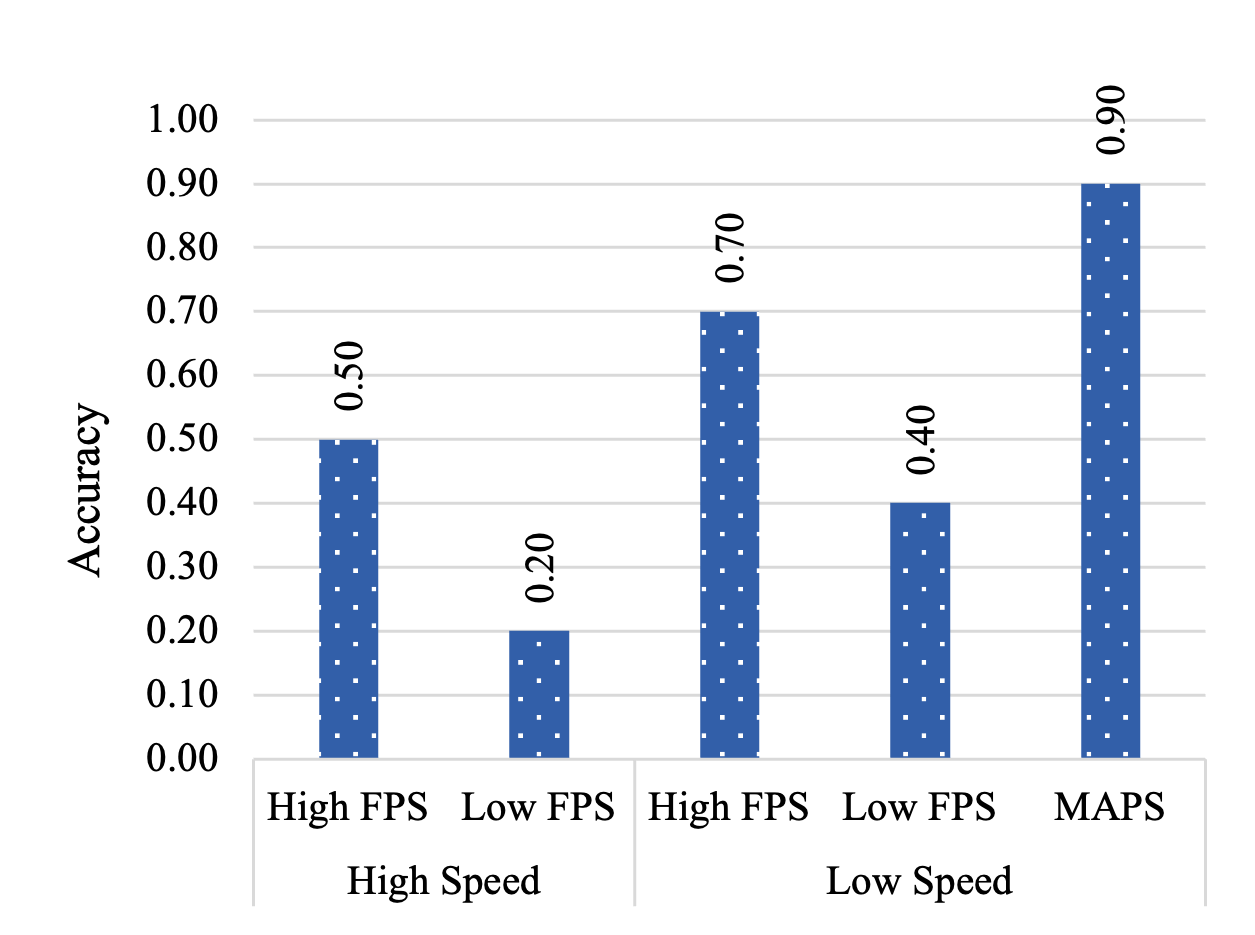}
	\caption{Accuracy of \proposedapp compared to baseline}
	\label{fig:ACC}
\end{figure}

To demonstrate the effectiveness of the proposed method, we compared it against four different baseline methods in our experiments: 1) High speed and High FPS  (Speed:90, FPS: 30), 2) Low speed and High FPS (Speed:70, FPS: 30), 3) High speed and Low FPS (Speed:90, FPS: 5), and 4) Low speed and Low FPS (Speed:70, FPS: 5). The experiments were conducted identically for each scenario, including the proposed method. The exact distance was considered for all tests.  For testing the scenarios, we considered driving five laps on the constructed route. We considered the robot's navigation accuracy and ability to stay on the designated path to measure reliability. As shown in Fig. \ref{fig:ACC}, the \proposedapp method achieved higher accuracy than the baseline scenarios. The \proposedapp method improved accuracy by 20\% compared to the best baseline scenario, which was Low speed and High FPS.

\begin{figure}[t]
	\centering
	\includegraphics[scale=0.45]{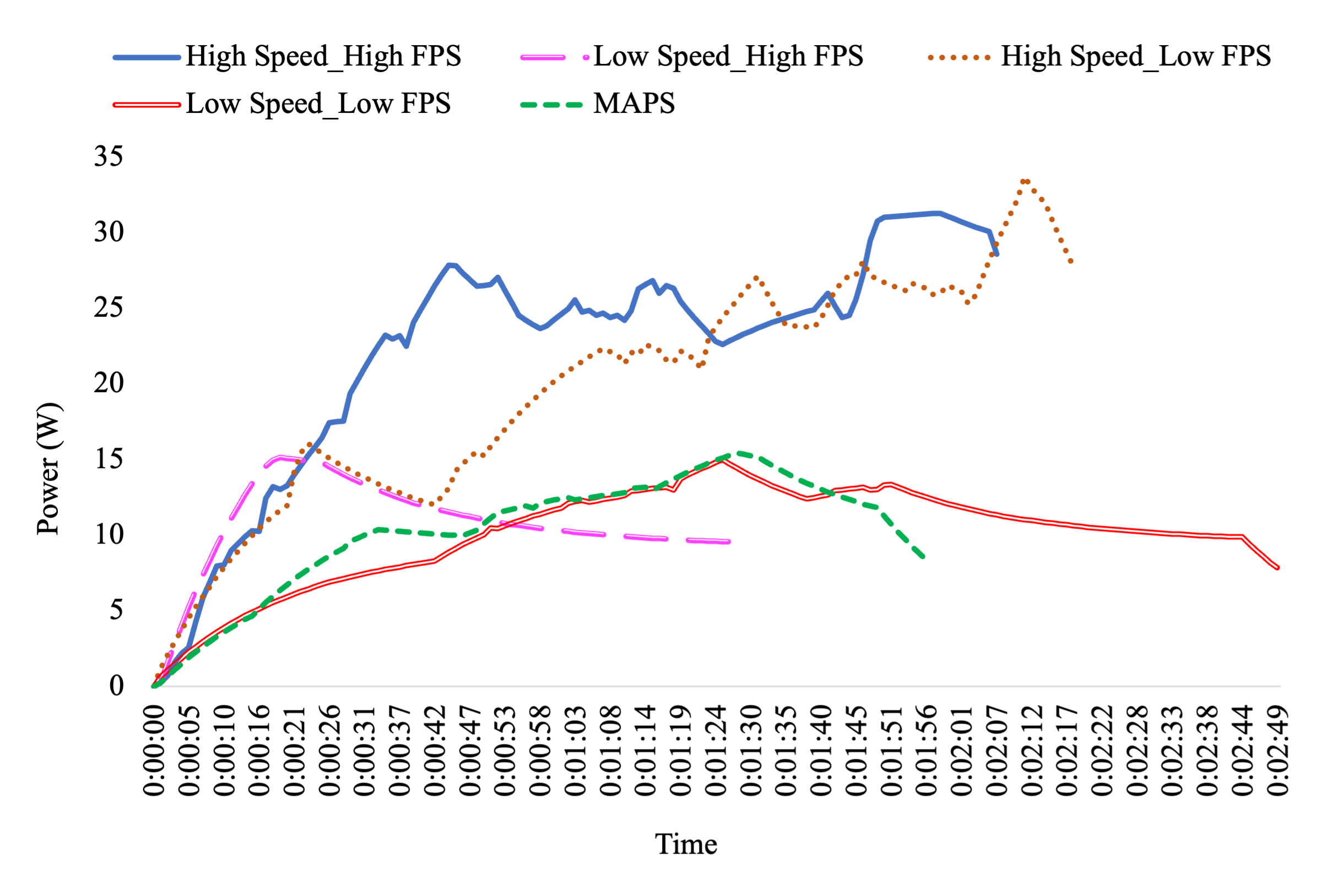}
	\caption{The amount of power consumed over time for all scenarios}
	\label{fig:PowerTime}
\end{figure}

\begin{figure}[t]
	\centering
	\includegraphics[width=\linewidth]{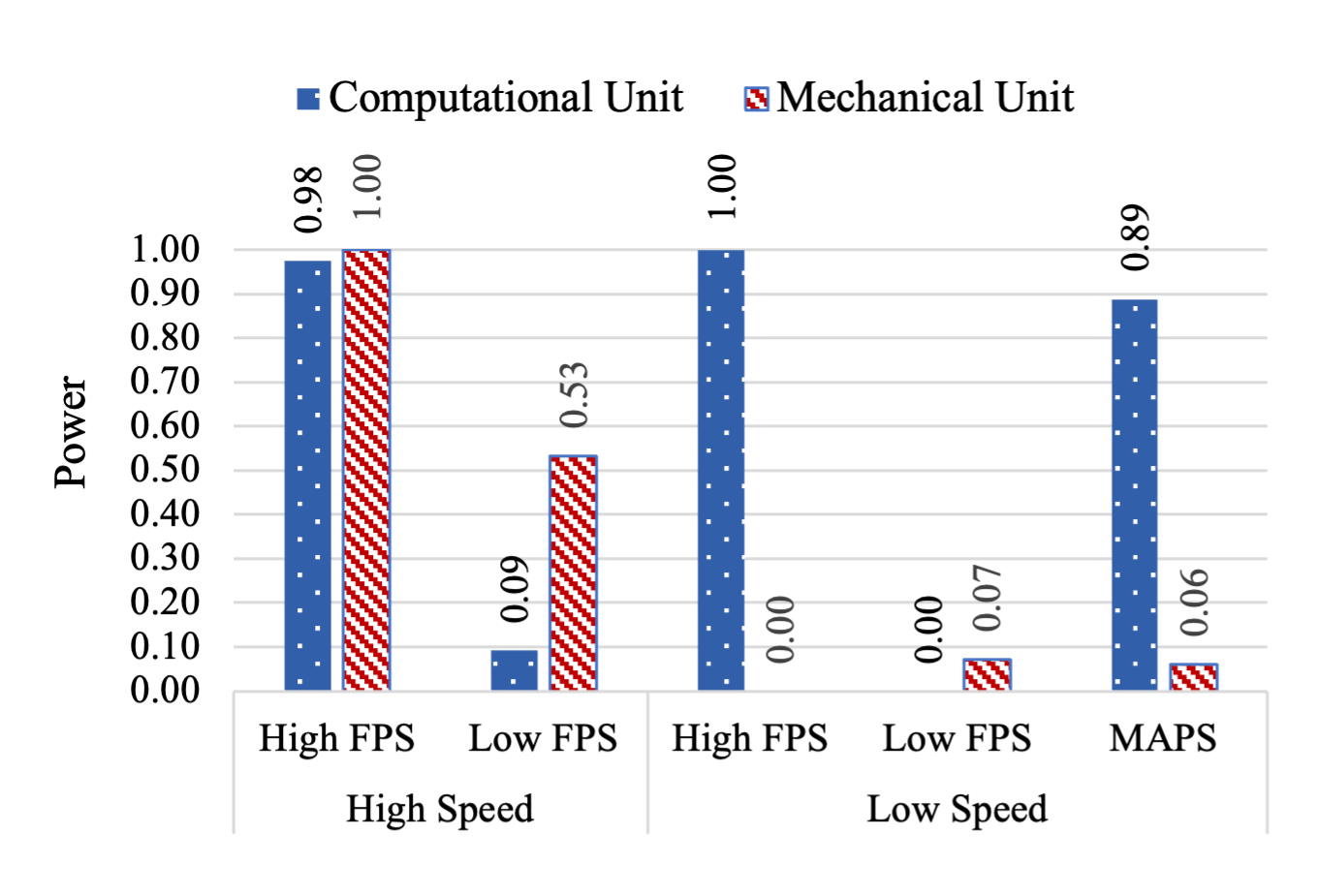}
	\caption{Power Consumption of Computational and Mechanical unit of \proposedapp compared to baseline (Normalized Number)}
	\label{fig:PowerSep}
\end{figure}

\begin{figure}[t]
	\centering
	\includegraphics[width=\linewidth]{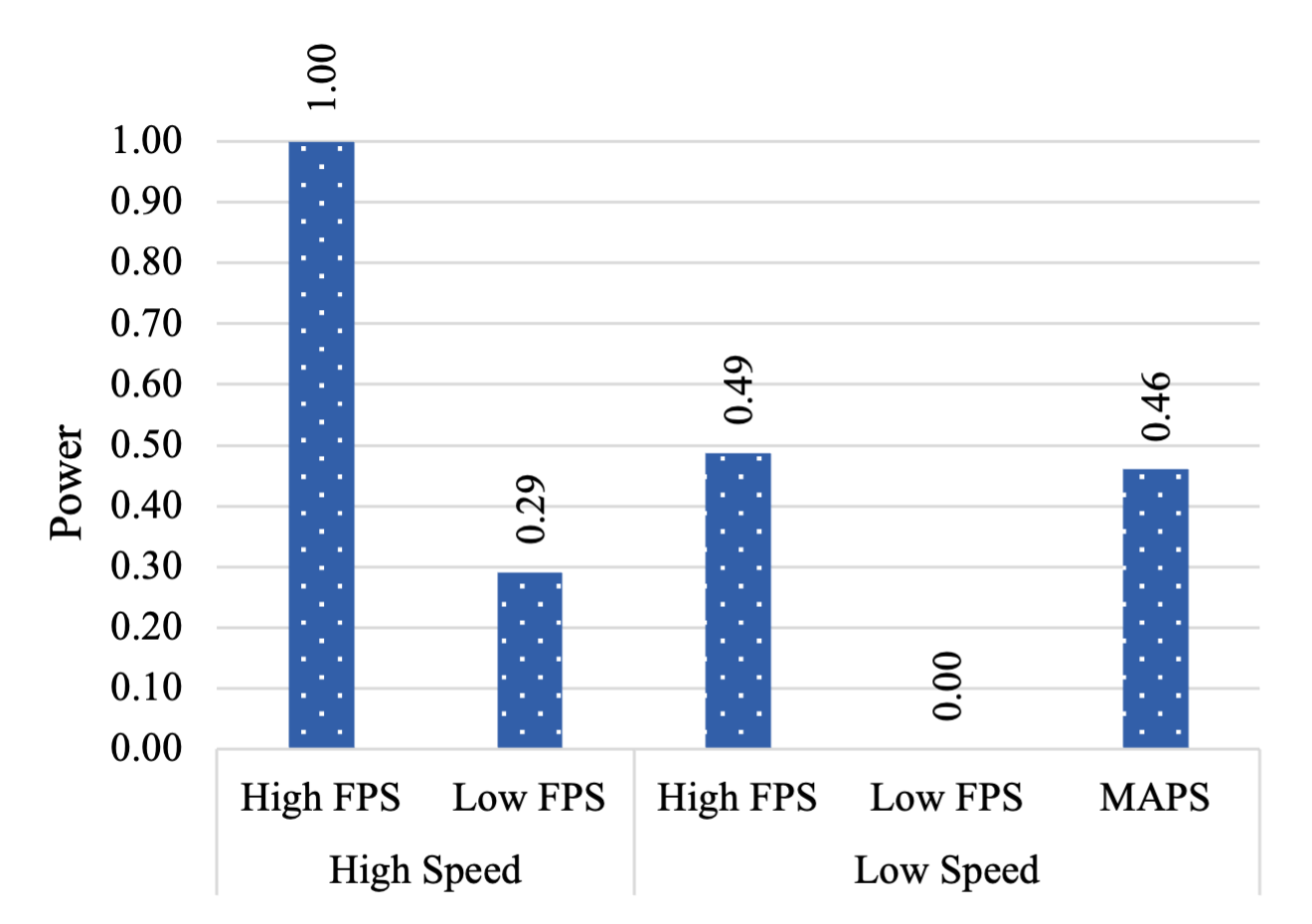}
	\caption{Total Power Consumption of \proposedapp compared to baseline (Normalized Number)}
	\label{fig:TotalEnergy}
\end{figure}

According to the experimental method mentioned, the power over time for each scenario can be observed in Fig. \ref{fig:PowerTime}. The measured power in Fig. \ref{fig:PowerTime} represents the sum of the computational and mechanical power for each scenario, shown separately. The proposed method completed the tests in a shorter duration compared to the High Speed High FPS, High Speed Low FPS, and Low Speed Low FPS scenarios. Additionally, it consumed less power compared to the High Speed High FPS and High Speed Low FPS scenarios.

Fig. \ref{fig:PowerSep} and \ref{fig:TotalEnergy} show the normalized power consumption of each unit and the total power consumption, respectively. The power consumption has been normalized based on the minimum and maximum power values. As shown in Figure \ref{fig:PowerSep}, the \proposedapp method does not have the lowest power consumption in the computational unit, but it managed to save 11\% power compared to the scenario with low speed and high FPS, which had the highest computational power consumption. In the mechanical unit, the \proposedapp method had the lowest power consumption compared to all scenarios. Additionally, as shown in Fig. \ref{fig:TotalEnergy}, the \proposedapp method performed well compared to the High speed, High FPS, Low speed, and High FPS scenarios, achieving 3\% and 54\% overall power savings, respectively. Therefore, it can be concluded that the \proposedapp method has demonstrated good performance in the trade-off between energy consumption and accuracy.
\section{Conclusion}
\label{conclusion}

The \proposedapp method presented in this paper demonstrates significant advancements in autonomous vehicle navigation and energy efficiency. Through rigorous testing and comparison against established baseline methods, the \proposedapp approach has shown superior navigation accuracy and energy savings performance. The integration of lane detection algorithms, OpenCV, and the GPIO library for motor control, along with efficient energy measurement techniques, has culminated in a robust system capable of optimizing autonomous vehicle operations.

A key highlight of the \proposedapp method is its ability to trade between computational and mechanical energy consumption while maintaining high navigation accuracy. The experimental results indicate that the \proposedapp method improved navigation accuracy by 20\% over the best-performing baseline scenario, which employed low speed and high frames per second (FPS) settings. This accuracy is critical for the reliable operation of autonomous vehicles, ensuring they can precisely navigate complex environments.

Moreover, the \proposedapp method achieved notable energy savings. While it did not achieve the lowest energy consumption in the computational unit, it still saved 11\% more energy than the highest energy-consuming baseline scenario. In the mechanical unit, the \proposedapp method outperformed all baseline scenarios in terms of energy efficiency. Overall, the \proposedapp method achieved up to 54\% total energy savings compared to other high-speed, high-FPS scenarios, underscoring its effectiveness in reducing energy consumption without compromising performance.

In conclusion, the \proposedapp method exemplifies a promising direction for future research and development in autonomous vehicle technology. This approach paves the way for more efficient and reliable autonomous systems by focusing on the trade-offs between energy consumption and navigation accuracy. Future work could explore further optimization techniques and broader applications to enhance the scalability and adaptability of the \proposedapp method across different autonomous vehicle platforms and environments.
\section{Appendix}
\label{Appendix}

\begin{figure}[t]
	\centering
	\includegraphics[scale=0.05]{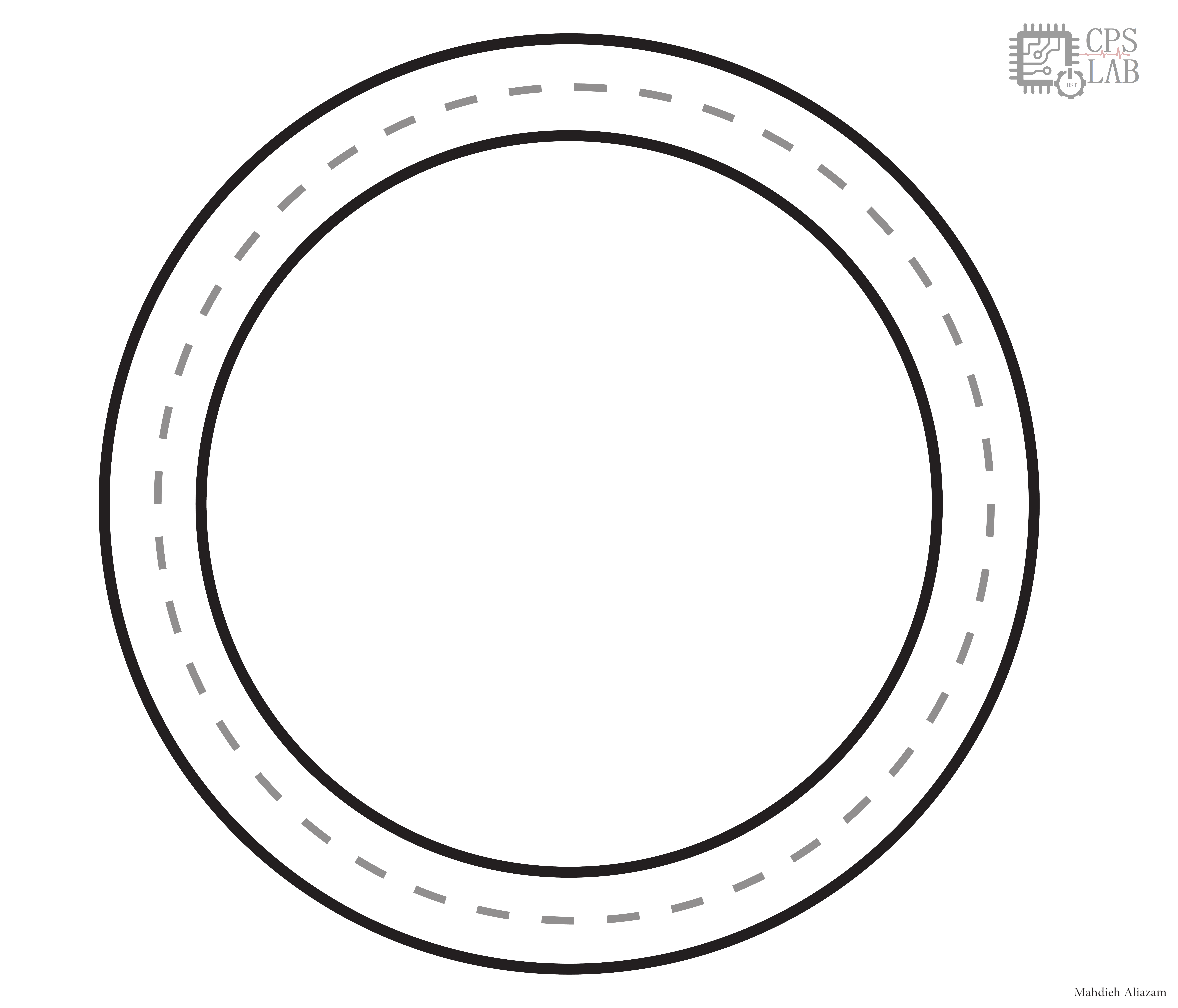}
	\caption{The proposed road path}
	\label{fig:Road}
\end{figure}

\textbf{Road image for conducting experiments:} The proposed road path is shown in Fig. \ref{fig:Road}.

Selected Prompts and Response of the LLM (GPT4o)

\textbf{Example:} Checking the overall movement path of the autonomous vehicle and getting the specific speed and FPS

\textbf{Prompt:} \emph{ We have a road with the given image specification. The robot can move at a speed between 70 and 90, measure images with the camera at 5 fps and 30 fps rates, and recognize its direction of movement to the right, left, and opposite. According to the specified route, tell us the appropriate speed and fps in different parts of the road.}

\textbf{ChatGPT-4o Response:} To determine the appropriate speed and frames per second (fps) rate for a robot navigating a circular road, we need to consider different parts of the route and how these factors affect the robot's performance.

Route Analysis
The image shows a circular road with both solid and dashed lines. These lines might indicate different zones or sections where speed adjustments are necessary:

1. Straight Sections: Likely on the edges of the circle.

2. Curved Sections: Throughout the circle, especially on the inner and outer paths.
Speed and FPS Settings

Straight Sections:
\begin{itemize}
    \item Speed: 90 (maximum speed)
    \item FPS: 5 fps (sufficient for detecting minor changes and staying on course)
\end{itemize}

Curved Sections:
\begin{itemize}
    \item Speed: 70-80 (reduce speed to handle turns effectively)
    \item FPS: 30 fps (higher fps to capture more frequent updates for better maneuvering)
\end{itemize}

Direction Recognition
The robot must recognize its movement direction (right, left, or opposite). This recognition is crucial for:
\begin{itemize}
    \item Making accurate turns.
    \item Adjusting speed appropriately.
    \item Staying within the designated lanes.
\end{itemize}

Speed and FPS Adjustment Strategy
\begin{itemize}
    \item Entering a Curve: Detect the curve and decrease speed to 70-80 while increasing fps to 30.
    \item Exiting a Curve: Once the curve ends, gradually increase speed back to 90 and decrease fps to 5.
    \item Straight Paths: Maintain the highest speed (90) and lower fps (5) for efficiency.
\end{itemize}

Practical Application

Based on this analysis, the robot should dynamically adjust its speed and fps based on its current position on the road:
\begin{itemize}
    \item Identify the current section (straight or curved).
    
    \item Adjust speed and fps accordingly:

     - Straight Path: Speed = 90, FPS = 5
     - Curved Path: Speed = 70-80, FPS = 30
    \item Utilize direction recognition** to make necessary adjustments when changing sections or lanes.
\end{itemize}

By implementing this strategy, the robot will navigate the circular road efficiently, maintaining optimal speed and accurate lane adherence.

\balance
\bibliographystyle{IEEEtran}
\bibliography{references}

\ifCLASSOPTIONcaptionsoff
  \newpage
\fi

\end{document}